# A simple en,ex marking rule for degenerate intersection points in 2D polygon clipping

Romeo Traian POPA[1], Emilia-Cerna MLADIN[1], Emil PETRESCU[1], Tudor PRISECARU[1]

**Abstract**
A simple en,ex rule to mark the intersection points of 2D input polygon contours separating the polygon interior from its exterior in the vicinity of the intersections is presented. Its form is close to the original Greiner & Hormann algorithm rule but encompasses degenerate intersections that are not self-intersections. It only uses local geometric information once the hand of the two input contours is known. The approach foundation is the distinction between two features of the studied intersections: the geometric intersection point and the assembling/concatenation point of the result contour/border. No special form of the intersection finding procedure is required.

## 1. Introduction

Boolean operations on 2D polygonal regions have a lot of important applications today, all stemming from the current widespread use of computers to create, handle and process graphical image.

Four algorithms have become in time, a classic presence in the field and are extremely popular: two created in the 70s, Sutherland & Hodgman [1] and Weiler & Atherton (W&A) [2] and the other two in the 90s, Vatti [3] and Greiner & Hormann (G&H) [4]. While the first is the simplest it can exclusively be used for convex polygons. The last two apply to concave polygons and polygons with holes or even self-intersecting. The original W&A could not approach self-intersecting polygons but Chakraborty [5] has quite recently made a proposal for such an extension.

G&H is certainly the simplest and most elegant of the last three algorithms but it can only correctly handle a very peculiar type of input contour intersections, the so--called non-degenerate intersections (where the intersection of the two input edges is a strictly interior point for both of them) . The solution proposed in the original G&H for handling degenerate intersections (points of an input contour that lie on the other input contour) is to "perturb" (slightly move) such configurations in order to turn them into non-degenerate situations. But once that is done, the result will no longer be unique (or said in another way, deterministic) as it will depend on the way the perturbations have been made.

---

[1] University POLITEHNICA of Bucharest, Romania

Kim & Kim (K&K) [6] proposed in 2006 an extension of the original G&H to deterministically handle degenerate intersections. According to their own statement, their solution does not cover the cases where such intersections are also self-intersecting points. The en,ex rule in their paper is more complicated and uses not simple, as in the original, but double en,ex flags that require non-local geometric information (not related to the intersection vertex in question only).
Foster & Overfelt made two attempts [7, 8] to simplify this aspect of the K&K approach but it was later discovered their solution did not cover all situations as they first thought .

We here present an en,ex rule that is simple and only uses for the marking process information related to a topological neighborhood (vicinity) of the intersection point. Our rule covers the same situations claimed by the K&K recipe, namely the two input polygons can be of the self-intersecting kind but the intersections of their contours must not be self-intersections (and the involved contour parts must be borders between polygon interior and exterior in the vicinity of any intersection point) . The rule does not require that the intersection test be of the form in the original G&H (where the "brute force" intersection approach is used).

The problem is only solved in this paper for the polygonal region intersection operation but finding the result polygonal planar region for the union and difference operations is very similar to the considerations presented and should be trivial.

Finding the result polygonal region is treated here in the traditional way, working not with the input regions themselves but with their oriented contours (each being in the "computer mind" an ordered list of polygon vertices). So the problem is viewed by us as trials made by the clipper polygon contour/border (by convention the clipper is "fixed"; meaning it is always the same polygon during the whole operation and its contour is drawn in red) to cross the subject polygon contour (which is always the same during the whole operation and drawn in black). Some trials will be successful and others will not.

Classifying all problematic types of trials (called by us non-trivial) that the original G&H ignores and selecting the 3 Rules which provide a correct en,ex marking/flagging in those cases is the content of Section 2 of the paper.
We here give, among others, an answer to the question : what color should be selected for the result contour in a red and black overlap situation where we have this option ? Obviously the two possible answers are : either red or black. But we think there are in principle at least two strategies to give the answer : an "asymmetric" option that we call the "fixed clipper" approach (and is the one followed in this paper) is to always choose the same color (here, we always choose black); and a different approach (which is presented in another paper that is going to be submitted to publication [9] ).

Section 3 deals with an unexpected simplification of the result of Section 2 obtained by adding an apparently unnecessary Rule 4. The now 4 Rules can be compressed in a sole, simple RULE, that uses just local geometric info about the intersection vertex to mark/flag.

In Section 4, we provide a visually very suggestive way of thinking about the analysed problem, by defining the so-called by us fundamental *on* intersections involved in the non-trivial trials (and assiging them en,ex marks). Then we simbolically represent the "component parts" of such trials as planar "lego" parts. The en,ex marks of such an ensemble (a non-trivial trial) can then be derived by an "addition operation" involving the marks of its ends (which are fundamental *on* intersections).

The Conclusions form Section 5 of this paper.

## 2. Trivial and non-trivial crossing trials by the clipper contour

Our task is now to deal with degenerate intersections and identify some procedure to corectly select the assembling points for the bicolor result border (which is made by concatenating red and black pieces that alternate and belong to the input polygon borders). We consider the intersection finding test is over at this moment; and the two vertex lists (red and black) have been enreached/completed (where a found intersection is a vertex present in the initial vertex list it has been accordingly flagged as an intersection/I vertex while where a found intersection is not an initial vertex, it has been inserted in the vertex list at its proper position and marked as an I vertex; every intersection point in the plane is actually a pair of linked vertices, a red I vertex and a black I vertex in respectively the red and black enreached list).

But before doing anything further, we decide right from the start that of the two input polygons, one is the clipper (and draw its contour in red) and the other is the subject (and draw its contour in black).

Two examples of degenerate intersection situations are presented in Fig. 1 and Fig. 2 (for visibility, the red contour parts overlapping the black contour have been drawn as very close parallels). We quickly realize the first situation is not as serious a problem as the second one. The key difference between the two is the first has only intersections consisting of one isolated (bicolor) point while in the second situation we have a series of consecutive intersection vertices (at least two) in the complete list. Why slightly changing the configuration to turn it into a non-degenerate one (as proposed in the G&H paper) is not a satisfactory solution is very suggestively illustrated in Figs. 2' and 2" (drawn following an idea about this aspect presented by Martinez & al [10]). The result polygonal region (the difference of the input regions in Fig. 2) differs with the way the changes/perturbations have been made.

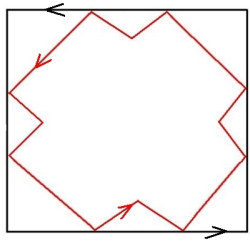

Fig. 1.   Isolated degenerate intersection points made by red and black contours

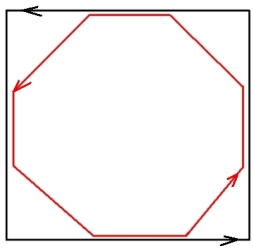

Fig. 2.   Degenerate intersections in the form of red and black contour overlaps

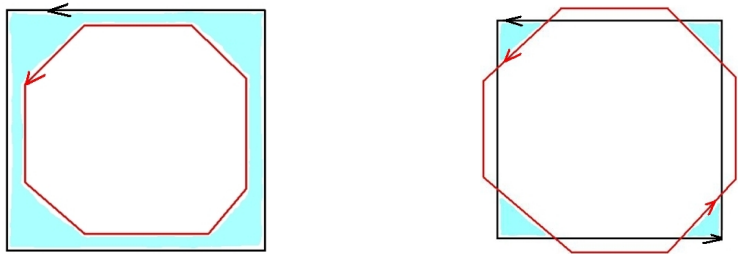

Figs. 2' and 2".   The difference (cyan) of the black and red regions depends on the way the red and black overlap is slightly changed in order to get rid of that degenerate situation

The new target is to closer examine such degenerate situations (we think of them as trials made by the red contour to cross the black one) where there is a series of at least two successive intersection vertices in both completed vertex lists, the red list and the black one. A red crossing trial begins either in the interior (*in* ) region of the black polygon or in its exterior (*out* ) region. A successful trial started in the *in* region ends in the *out*  region and a successful trial started in the *out*  region ends in the *in*  region. Fortunately, the total number of trials is finite and very small so we can look at all of them case by case.
So all possible non-trivial situations are illustrated in Fig. 3 (where the interior of both the red and the black polygon is always considered on the left of them; in other words both contours are of left hand). For the  convenience of drawing, a part of the contour is not shown as a sequence of line segments but as a curved line. The name of the trials is given from the clipper perspective (they are clipper's trials) : *in,out on_con* means a finally successful trial of the clipper (red) contour to cross the subject (black) contour when coming from the inside of subject (*in* ) . So when the crossing trial is finished, the red contour is outside (*out* ) of the "black" polygonal region. *on* refers to the fact the crossing trial is a non-trivial one; such a trial has a red and black contour overlap (which is reflected in the red and black completed lists by a series of at least two successive vertices flagged as I vertices). *con*  tells us the red and black contour parts that overlap, point the same way (illustrated in the drawing by the arrow tips). When their flows are opposite, we use *on_opp* instead of *on_con* in the name of the trial. The trial is thought of from the clipper perspective, so it follows the flow of the red contour : its first intersection point of the series is denoted A in the drawing and its last, B.

We can very easy decide, by visually examining every of the 8 cases in Fig. 3, how to correctly en,ex mark/flag an intersection point belonging to the red contour such that the result contour (made of concatenated red and black pieces that alternate) is the one we are looking for (the result here is that of the intersection operation, namely the polygonal region common to the red interior and the black interior; in Fig. 3, that intersection region is colored cyan close to its border).

But first we recall the significance of the red en and ex flags in the original G&H : en is a (bicolor) intersection point where the red contour enters the black interior/inside (so from that red vertex on, the vertices in the red list belong to the result/intersection bicolor contour); ex is a (bicolor) intersection point where the red contour exits the black interior/inside (so from that ex red vertex, further on in the red list, the vertices do not belong to the result/intersection bicolor contour).
The new situation we face now (that does not have a place in the above en and ex mark/flag definitions as they only consider non-degenerate intersections) is a red and black overlap. In these new cases (and all possibilities are the 8 cases in Fig. 3), we try when building/assembling the red and black result contour, to :

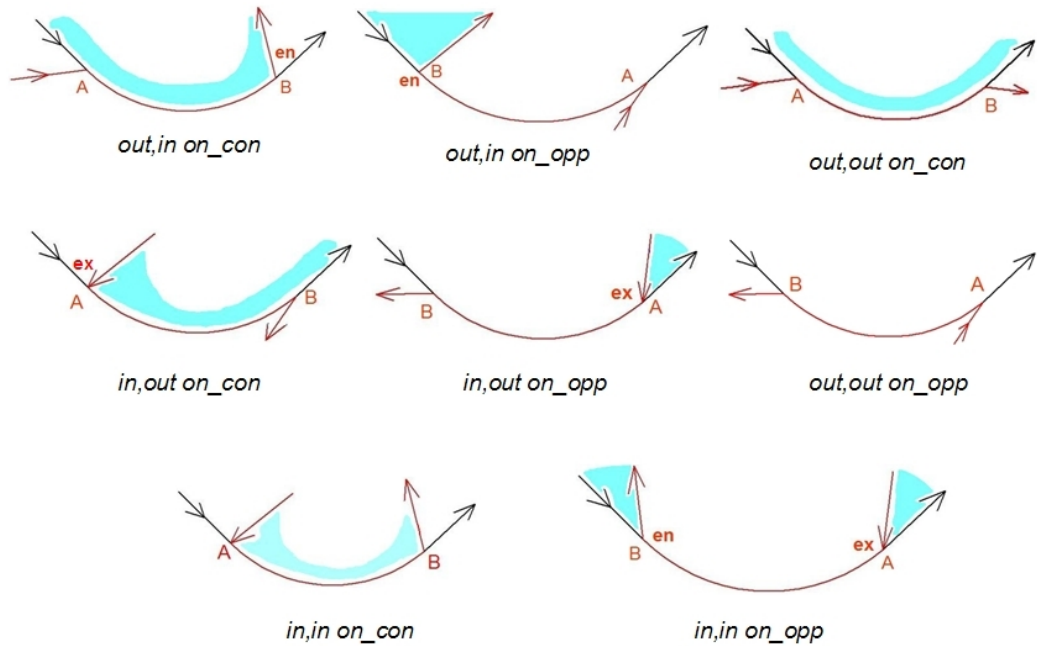

Fig. 3   All possible non-trivial trials by the left hand red contour to cross the left hand black contour
(the intersection region of the two polygons is colored cyan)

make the minimum number of concatenations (why should we change the color of the result contour if we do not have to ?) and not include in the result polygonal region, "degenerate" surfaces (surfaces whose area is zero, that consist of overlapped red and black contour parts having opposite flows).

An example of how such degenerate surfaces can be created when the concatenation point (flagged as en or ex in the red list) is not adequately chosen is illustrated in Fig. 4a and 4b. If of all successive I (red) vertices belonging to the series A to B in Fig. 4a , we choose a red vertex that is different from B, be it C, then the intersection region of the red and black polygons will look like in Fig. 4b, having a non-zero area surface plus an undesired attachment on its right : a degenerate surface that is an overlap of the red contour part from the vertex C to B and the black contour part from the vertex B to C (where the black I vertices B and C are respectively linked to the red vertices B and C).

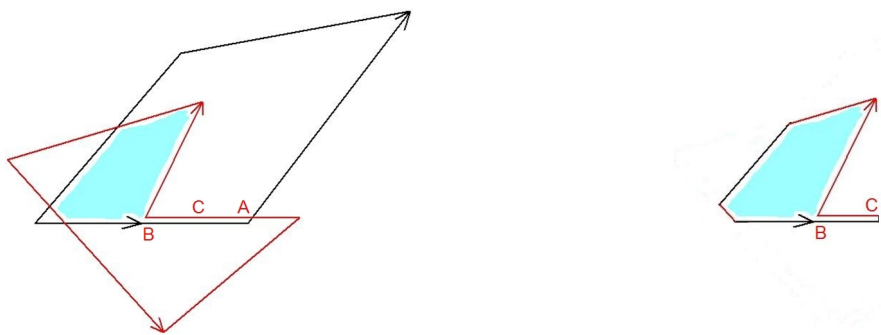

Figs. 4a and 4b.   Inadequately selecting C as concatenation point instead of B, adds a zero area
surface (a red and black contour overlap) to the polygon intersection region
which is drawn in the right image (the result interior is colored cyan)

By following the above two principles and selecting for simplicity the same rule for the *in,out on_con* and *out,in on_con* cases as for their *on_opp* correspondents, we visually decide the correct en and ex points in all 8 cases in Fig. 3; the selected points and their en,ex flags are shown in the image. Now if we express in words the results of that selection, we have 3 Rules for the clipper:
Rule1: for an *in,out* successful crossing trial, the concatenation point (an ex vertex) is the first
 (red) vertex of the trial
Rule 2: for an *out,in* successful crossing trial, the concatenation point (an en vertex) is the last
 (red) vertex of the trial.
Rule 3: for an *in,in on_opp* crossing trial (thus an unsuccsessful trial), the concatenation points are
 both ends of the trial (the start of it is an ex vertex and the end of it is an en vertex).

The three rules stated above cover all circumstances where a concatenation point of the result bicolor contour has to be defined, see Fig. 3.

Besides the non-trivial crossing trials discussed so far, there also are trivial crossing trials, consisting of just one. isolated, intersection/I point/vertex. These too can be successful or unsuccessful; and those successful where the intersection point is strictly interior to the red and black arrows/oriented edges are the non-degenerate intersections the original G&H so elegantly deals with, see Fig. 5a.
However, after enreaching/completing the initial vertex lists, all one vertex crossing trials are "geometrically the same", see Fig. 5b: a set of 4 arrows, 2 red (one entering the intersection and one exiting it) and 2 black (one entering the intersection and one exiting it). If the location of the black and red interior regions is known (that is if we know the hand of the red and black contours in the vicinity of that intersection point. either left or right), then the type of the respective one vertex trial (*in,in* , *in,out* , *out,in* or *out,out* ) can be obtained by using the cross product of that intersection's arrows (detailed comments on the issue in [9] and [11]). In the case of an *in,out* trial the red intersection vertex is of course flagged ex; it is flagged en for an *out,in* trial.

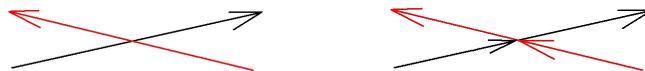

Figs. 5a and 5b.  A non-degenerate intersection, properly handled by the original G&H (left); the 4
 arrows of it and any one-vertex trial after vertex lists have been completed (right)

There is one more type of intersection point : the *on,on* case (details on classification of red-and-black-edge intersections are given in [9]). But the *on,on* type is "neutral" regarding the crossing process, it neither advances nor draws back the clipper (red) contour; such intersection points cannot be ends of a trial but only mid parts of it and they will not receive any en or ex flags, the same as the one (I) vertex *in,in* and *out,out* trials.

## 3. A simple en,ex rule for the fixed clipper contour

Let us decide to break the principle of the minimum number of concatenation points/vertices (that we used when we derived the 3 Rules for the clipper) in the following way : we add an apparently unnecessary Rule 4, in the case of the *in,in on_con* trial (see Fig. 3), flagging both ends of it as in the case of the *in,in on_opp* trial (the start of the trial will be ex and the end of it en). In this way, we have given precedence to the primciple of the "fixed clipper" (the red polygon) : wherever there is a red and black overlap that belongs to the result contour, we build the result contour using the black color.

What we have got now, is the possibility to flag an intersection/I vertex based solely on geometrical info related to its vicinity. In other words, we can assign en or ex flags to some "fundamental intersection" types the non-trivial crossing trials consists of (see next section). Summarizing now the decisions currently applied to both the non-trivial and the trivial crossing trials of the clipper, the following simple RULE applies :
- intersection points of the type *in,out* and *in,on* are flagged ex
- intersection points of the type *out,in* and *on,in* are flagged en
- all other red intersection/I vertices are not en,ex marked/flagged

Altough the hand is left for both the red contour and the black contour in Fig, 3, the above RULE for the clipper (the red polygon) holds for every combination of their hands. We stress that is not always true for the subject which is drawn black here. In some situations it is, in others it is not (details are provided in [11]).
Said otherwise, it is compulsory in the "fixed clipper" approach to first set the en,ex flags (of the red list). After that, the en,ex flags of the black list are derived according to the following "asymmetric" procedure: a black I vertex linked to an en red vertex will be ex if the red and black contours have the same hand in the respective intersection point and will be en if viceversa; a black I vertex linked to an ex red vertex will be en if the red and black contours have the same hand in that intersection point and will be ex if viceversa.

In case the wanted result is the union of the red and black input polygons, the RULE is the following (of course this time, the border of the result polygon will be the union of the contour parts, either red or black, that are outside the other polygon) :
- intersection points of the type *in,out* and *on,out are* labeled en
- intersection points of the type *out,in* and out,on are labeled ex
- all the other intersection/I vertices are not en,ex labeled

All the above assume we know the hand of the red and black contours in the vicinity of the analyzed intersection (whose en,ex flag will be set where the case). If the two input contours are simple polygons, for each of them the hand is the same along the entire length. In such a case, to find out the unique hand of the whole contour is enough to perform once a point in polygon test for a non-intersection point of the contour (not necessarily a vertex) in the vicinity of the intersection in question. Because the tested non-intersection point is very close to the intersection (but different from it), its location mark relative to the other polygon, either *in* or *out*, will be the same *in* or *out* necessary to define the type of that intersection (that point will belong to one of the 2 red arrows or to one of the 2 black arrows of that I vertex). From the two location marks calculated this way (one for the red contour and one for the black contour), the second red location mark for the intersection in question (the one related to the other red arrow), can be derived by using the cross product of its 4 (2 red and 2 black) arrows.

### 4. A suggestive graphic description using planar "lego" parts

The RULE just stated in Section 3 gives for every red I vertex its en or ex flag where that is the case (some I vertices, like for instance *out,out* , or *out,on* or *in,in*  will receive no en or ex flag). The importance of the rule is not just its compactness but the fact that once the hand of both contours in the vicinity of the intersection is known, it only needs local information : the two location marks relative to the black interior, one for each of the 2 red arrows of the analysed intersection .

We will call "fundamental *on* intersections" , all intersections points appearing as ends of non-trivial crossing trials by the clipper. Their types are 8 in number and are represented, together with their en or ex flags, in Fig. 6. Their name shows the presence of an *on* half in them (that part of them where one red and one black arrow overlap). If we try to build/assemble such crossing trials, the building parts/pieces we need are their ends and any number (zero included) of "neutral" connectors (which are, for a given trial, identical *on,on* intersection points/vertices). There are two types of the neutral connectors : the *on_con* type where the overlapping red and black arrows/contours flow the same way and the *on_opp* type where the two overlapping arrows have opposite flows. So every fundamental *on* intersection has one half that is half of a neutral (that is *on,on* ) connector.

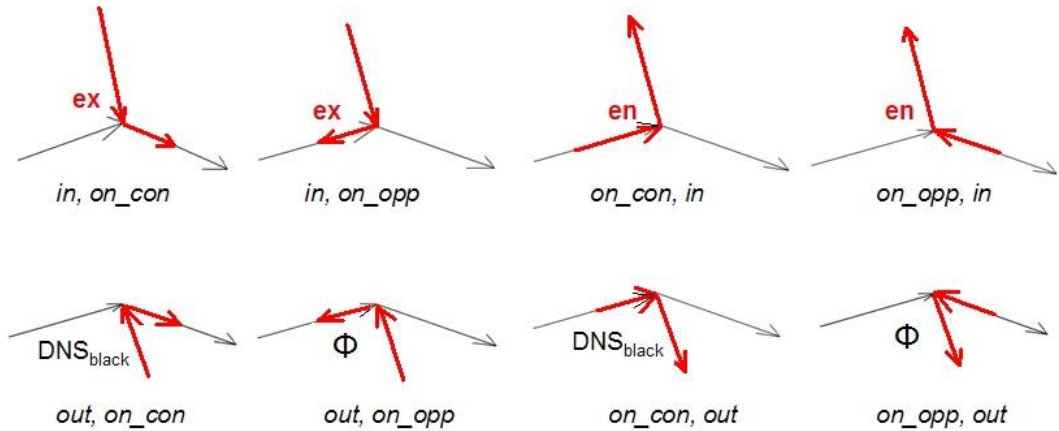

Figs. 6 The fundamental *on* intersections of the left red contour and their flagging status for the intersection boolean operation (only the intersection points with red labels are flagged in the algorithm) : Φ just signals for the reader there is no result contour there and $DNS_{black}$ means Do Not Switch black color for the result contour (both black arrows belong to the result contour; here, the black contour also is of left hand, having the black polygon interior to its left)

In conclusion, we can suggestively visualize any non-trivial red crossing trial as built by using the enumerated "lego" parts : two ends, plus, if the trial is a series of more than 2 I vertices, a finite number of *on,on*  connectors of the same type.
Although in the drawing, the red contour can flow in any direction (from right to left or from left to right and so on), in the symbolic lego representation of a non-trivial trial, the contour always flows from left to right like ordinary text.

Table 1 displays just 2 of the 8 types of non-trivial crossing trials of the clipper (red) contour, as assembles of certain end parts and possibly, a finite number of neutral connectors in between : the

*in,out on_con* trial and the *in,in on_opp* trial. It is obvious the continuity of the flow of the red and black contours between the ends of the trial requires a certain match between the *on* halves of the end parts and between them and the neutral connectors in between : all have to be of the same type, either *on_con* or *on_opp* .

If the en or ex "mark" assigned to an end part is written a little differently, as a pair of components, either en or ex and 0 (zero), then the "mark" of the whole ensemble (the non-trivial trial) is obtained as the "addition" of the pair marks of its ends. 0 signifies the red part of the overlap (the on part of the red contour in the vicinity of that intersection point) is not in the result contour. For instance, for the *in,out on_con* trial and *in,in on_opp* trial we write respectively:

$$(ex,0)+(0,0) = (ex,0)$$
$$(ex,0)+(0,en) = (ex,en)$$

It is obvious that if we extended this symbolic representation to the trivial trials, they would lack the *on,on* connector formed for a two vertex trial by the two *on* halves of its ends (see the column Crossing trial type in Tabel 1).

Tabel 1. Two non-trivial red trials built of lego parts : trial ends and neutral connectors

|  | No | Trial ends | | Crossing trial type | Between-ends connector type |
|---|---|---|---|---|---|
|  |  | First end/vertex | Last end/vertex |  |  |
| Name | 1 | *in, on_con* | *on_con, out* | *in,out on_con* | *on_con* |
| Lego part (symbol) |  |  |  |  |  |
| Concatenation flag(s) |  | ex |  | ex |  |
| Flag position |  |  |  | first vertex |  |
| Name | 2 | *in, on_opp* | *on_opp, in* | *in,in on_opp* | *on_opp* |
| Lego part (symbol) |  |  |  |  |  |
| Concatenation flag(s) |  | ex | en | a pair : ex, en |  |
| Flag position |  |  |  | first and last vertex |  |

## 5. Conclusions

We have here given a simple rule to flag as en or ex points, the intersection vertices of the clipper polygon; a rule that extends the en,ex rule of the original Greiner & Hormann (G&H) algorithm [4] to degenerate intersections in situations with self-intersecting input contours having intersection points that are not self-intersections. The presented procedure has an "asymmetry" feature (it always selects the subject contour not the clipper contour when building the result contour of input contour pieces has such an option); and we have called it because of that, the "fixed clipper" approach. Our approach is simpler then the G&H extension shown in K&K [6] and uses only local geometric information (related to the intersection vertex to flag only) if the hand of the input contours in the vicinity of the respective intersection point is known. The form of the intersection finding test does not have to be the one used in the original G&H.

## Acknowledgements

The first author gratefully acknowledges the donation received from S.C. IWMS Consultant SRL that gave him the opportunity to extend the time for this research by two months.